\documentstyle[10pt,epsf]{article}

\catcode`\@=11
\long\def\@makefntext#1{ 
\protect\noindent \hbox to 3.2pt {\hskip-.9pt
$^{{\ninerm\@thefnmark}}$\hfil}#1\hfill} 

 \def\@makefnmark{\hbox to 0pt{$^{\@thefnmark}$\hss}}  

\def\ps@myheadings{\let\@mkboth\@gobbletwo
\def\@oddhead{\hbox{} 
\rightmark\hfil\ninerm\thepage}
\def\@oddfoot{}\def\@evenhead{\ninerm\thepage\hfil 
\leftmark\hbox{}}\def\@evenfoot{}
\def\sectionmark##1{}\def\subsectionmark##1{}}

\begin{document}


\newcounter{sectionc}\newcounter{subsectionc}\newcounter{subsubsectionc}
\renewcommand{\section}[1] {\vspace{0.6cm}\addtocounter{sectionc}{1}
\setcounter{subsectionc}{0}\setcounter{subsubsectionc}{0}\noindent
	{\bf\thesectionc. #1}\par\vspace{0.4cm}}
\renewcommand{\subsection}[1] {\vspace{0.6cm}\addtocounter{subsectionc}{1}
	\setcounter{subsubsectionc}{0}\noindent
	{\it\thesectionc.\thesubsectionc. #1}\par\vspace{0.4cm}}
\renewcommand{\subsubsection}[1] {\vspace{0.6cm}\addtocounter{subsubsectionc}{1}
	\noindent {\rm\thesectionc.\thesubsectionc.\thesubsubsectionc.
	#1}\par\vspace{0.4cm}}
\newcommand{\nonumsection}[1] {\vspace{0.6cm}\noindent{\bf #1}
	\par\vspace{0.4cm}}

\newcounter{appendixc}
\newcounter{subappendixc}[appendixc]
\newcounter{subsubappendixc}[subappendixc]
\renewcommand{\thesubappendixc}{\Alph{appendixc}.\arabic{subappendixc}}
\renewcommand{\thesubsubappendixc}
	{\Alph{appendixc}.\arabic{subappendixc}.\arabic{subsubappendixc}}

\renewcommand{\appendix}[1] {\vspace{0.6cm}
        \refstepcounter{appendixc}
        \setcounter{figure}{0}
        \setcounter{table}{0}
        \setcounter{equation}{0}
        \renewcommand{\thefigure}{\Alph{appendixc}.\arabic{figure}}
        \renewcommand{\thetable}{\Alph{appendixc}.\arabic{table}}
        \renewcommand{\theappendixc}{\Alph{appendixc}}
        \renewcommand{\theequation}{\Alph{appendixc}.\arabic{equation}}
        \noindent{\bf Appendix \theappendixc #1}\par\vspace{0.4cm}}
\newcommand{\subappendix}[1] {\vspace{0.6cm}
        \refstepcounter{subappendixc}
        \noindent{\bf Appendix \thesubappendixc. #1}\par\vspace{0.4cm}}
\newcommand{\subsubappendix}[1] {\vspace{0.6cm}
        \refstepcounter{subsubappendixc}
        \noindent{\it Appendix \thesubsubappendixc. #1}
	\par\vspace{0.4cm}}

\def\abstracts#1{{
	\centering{\begin{minipage}{25pc}\tenrm\baselineskip=12pt\noindent
	\centerline{\tenrm ABSTRACT}\vspace{0.3cm}
	\parindent=0pt #1
	\end{minipage} }\par}}

\newcommand{\bibit}{\it}
\newcommand{\bibbf}{\bf}
\renewenvironment{thebibliography}[1]
	{\begin{list}{\arabic{enumi}.}
	{\usecounter{enumi}\setlength{\parsep}{0pt}
\setlength{\leftmargin 1.25cm}{\rightmargin 0pt}
	 \setlength{\itemsep}{0pt} \settowidth
	{\labelwidth}{#1.}\sloppy}}{\end{list}}

\topsep=0in\parsep=0in\itemsep=0in
\parindent=1.5pc

\newcounter{itemlistc}
\newcounter{romanlistc}
\newcounter{alphlistc}
\newcounter{arabiclistc}
\newenvironment{itemlist}
    	{\setcounter{itemlistc}{0}
	 \begin{list}{$\bullet$}
	{\usecounter{itemlistc}
	 \setlength{\parsep}{0pt}
	 \setlength{\itemsep}{0pt}}}{\end{list}}

\newenvironment{romanlist}
	{\setcounter{romanlistc}{0}
	 \begin{list}{$($\roman{romanlistc}$)$}
	{\usecounter{romanlistc}
	 \setlength{\parsep}{0pt}
	 \setlength{\itemsep}{0pt}}}{\end{list}}

\newenvironment{alphlist}
	{\setcounter{alphlistc}{0}
	 \begin{list}{$($\alph{alphlistc}$)$}
	{\usecounter{alphlistc}
	 \setlength{\parsep}{0pt}
	 \setlength{\itemsep}{0pt}}}{\end{list}}

\newenvironment{arabiclist}
	{\setcounter{arabiclistc}{0}
	 \begin{list}{\arabic{arabiclistc}}
	{\usecounter{arabiclistc}
	 \setlength{\parsep}{0pt}
	 \setlength{\itemsep}{0pt}}}{\end{list}}

\newcommand{\fcaption}[1]{
        \refstepcounter{figure}
        \setbox\@tempboxa = \hbox{\footnotesize Fig.~\thefigure. #1}
        \ifdim \wd\@tempboxa > 4.5in
           {\begin{center}
        \parbox{4.5in}{\footnotesize\baselineskip=10pt Fig.~\thefigure. #1 }
            \end{center}}
        \else
             {\begin{center}
             {\footnotesize Fig.~\thefigure. #1}
              \end{center}}
        \fi}

\newcommand{\tcaption}[1]{
        \refstepcounter{table}
        \setbox\@tempboxa = \hbox{\tenrm Table~\thetable. #1}
        \ifdim \wd\@tempboxa > 6in
           {\begin{center}
        \parbox{6in}{\tenrm\baselineskip=12pt Table~\thetable. #1 }
            \end{center}}
        \else
             {\begin{center}
             {\tenrm Table~\thetable. #1}
              \end{center}}
        \fi}

\def\@citex[#1]#2{\if@filesw\immediate\write\@auxout
	{\string\citation{#2}}\fi
\def\@citea{}\@cite{\@for\@citeb:=#2\do
	{\@citea\def\@citea{,}\@ifundefined
	{b@\@citeb}{{\bf ?}\@warning
	{Citation `\@citeb' on page \thepage \space undefined}}
	{\csname b@\@citeb\endcsname}}}{#1}}

\newif\if@cghi
\def\cite{\@cghitrue\@ifnextchar [{\@tempswatrue
	\@citex}{\@tempswafalse\@citex[]}}
\def\citelow{\@cghifalse\@ifnextchar [{\@tempswatrue
	\@citex}{\@tempswafalse\@citex[]}}
\def\@cite#1#2{{$\null^{#1}$\if@tempswa\typeout
	{IJCGA warning: optional citation argument
	ignored: `#2'} \fi}}
\newcommand{\citeup}{\cite}


\font\twelvebf=cmbx10 scaled\magstep 1
\font\twelverm=cmr10 scaled\magstep 1
\font\twelveit=cmti10 scaled\magstep 1
\font\elevenbf=cmbx10 scaled\magstephalf
\font\elevenrm=cmr10 scaled\magstephalf
\font\elevenit=cmti10 scaled\magstephalf
\font\tenbf=cmbx10
\font\tenrm=cmr10
\font\tenit=cmti10
\font\ninebf=cmbx9
\font\ninerm=cmr9
\font\nineit=cmti9
\font\eightbf=cmbx8
\font\eightrm=cmr8
\font\eightit=cmti8
\centerline{\tenbf SCALING REGION MESON PHOTOPRODUCTION }
\baselineskip=16pt
\centerline{\tenbf AND ELASTIC FORM FACTORS OF HADRONS$^1$}
\vspace{0.5cm}
\centerline{\tenrm Andrei V. Afanasev$^2$}
\baselineskip=13pt
\centerline{\tenit North Carolina Central University, Durham, NC 27707}
\baselineskip=12pt
\centerline{\tenit and}
\centerline{\tenit Jefferson Lab, Newport News, VA 23606}
\vspace{0.6cm}
\abstracts{First, I define generalized Bloom-Gilman duality and Bjorken-like scaling for
inclusive photoproduction of pions, relating Quark-Parton Model description in the 
deep-inelastic region to the properties of exclusive resonance excitation. 
Secondly, connection between inclusive
and exclusive processes is established in the formalism of Nonforward 
Parton Densities used here to predict deviations from dimensional scaling for a variety
of observables: nucleon and pion elastic form factors, and pion Compton scattering.}
\footnotetext{$^1$ Based on two invited talks at the Workshop `Jefferson Lab Physics and Instrumentation with
6-12 GeV Beams', Newport News, VA, June 15-18, 1998}
\footnotetext{$^2$ On leave from Kharkov Institute of Physics and Technology, Kharkov, Ukraine}
\vfil
\rm\baselineskip=12pt
\section{Extending Bloom-Gilman Duality to Meson Photoproduction}
For inclusive lepton nucleon scattering, $N(l,l')X$, 
Bloom-Gilman duality is a relation between the deep-inelastic scattering 
region and the nucleon resonance region.\cite{bloom} It may be formulated
as follows: At large $Q^2$, leptoproduction cross sections of hadronic states 
with a fixed invariant mass may be adequately described in two physically equivalent ways,
a) as an integral of structure functions continued from the scaling region and
b) as a sum of squared form factors of nucleon resonance excitation. In other words,
scaling structure functions on average give a true
curve of resonance excitation, and this relation remains valid as $Q^2$ increases.
Duality shows that the fundamental single quark QCD process is decisive not only in 
the deep-inelastic regime, but also in 
setting the scale of the reaction in the resonance region. 

Let us extend duality studies to the process of inclusive pion (or other meson)
photoproduction, $\gamma+N\to\pi+X$, where the hadronic state $X$ may be either 
within or above the resonance region.  In the latter case and high transverse momenta of
the produced pion, one may apply a Quark-Parton Model for the single quark process, 
see Ref.\cite{acw} and references therein. For high enough photon energies, isolated pions directly 
produced via the mechanism of Fig.1  dominate as one approaches the 
kinematic upper limit, i.e., the resonance region.

\begin{figure}[h]
\let\picnaturalsize=N
\def\picsize{1in}
\def\picfilenamea{direct.eps}
\ifx\nopictures Y\else{\ifx\epsfloaded Y\else\input epsf \fi
\let\epsfloaded=Y
\centerline{
\ifx\picnaturalsize N\epsfxsize \picsize\fi \epsfbox{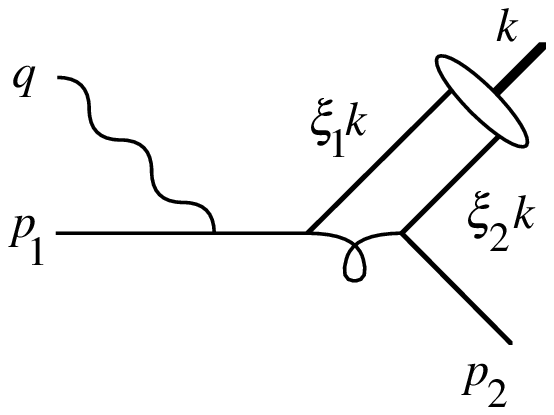}}}
\fi
\fcaption{Direct pion photoproduction off a quark via gluon exchange. There are
four diagrams total, corresponding to the four quark lines the photon 
may couple to.}
\end{figure}

For this process, one may define
a scaling variable $x$ in terms of the Mandelstam variables,
\begin{equation}
x= {-t\over s+u-2 m_N^2},\ \ 0\leq x\leq 1.
\end{equation}  
This new variable is a precise analog of $x_{Bjorken}$ in lepton scattering, i.e., 
a fraction of the target momentum carried by the struck quark. The value $x=1$ corresponds to
the exclusive limit of pion photoproduction ($m_X=m_N)$. The modified scaling variable, 
a necessary step in derivation of duality, is defined by analogy with Bloom's and Gilman's
proposal, with $-t$ replacing $Q^2$:
\begin{equation}
\omega '={1\over x'}={1\over x}+{m_N^2\over -t}.
\end{equation}
The duality relation then may be written as an $\omega'$-integral of the
differential cross section $d\sigma/dxdt (\gamma N\to\pi X)$ and in
the region of the direct process (Fig.1) dominance reads 
\begin{eqnarray}
\int_{1-{m_N^2\over t}}^{1-{M^2\over t}} d\omega '\sum_q G_{q/N}(\omega ') 
{d\sigma(\gamma q\to\pi q')\over dt}\propto\sum_R {d\sigma\over dt}(\gamma+N\to\pi+R)
\end{eqnarray}
Summation in the r.h.s. of Eq.(3) is done over all resonances $R$ with masses $m_R\leq M$,
with the nucleon final state included.
If the parton distribution function of the nucleon is $G_{q/N}(\omega ')\approx (\omega'-1)^3$
at $\omega'\to 1$ and the subprocess $\gamma q\to\pi q'$ cross section is determined 
by the one-gluon exchange mechanism of Fig.1, then duality (3) requires that the resonance excitation 
cross section $d\sigma/dt(\gamma+N\to\pi+R)\propto 1/s^7$ at fixed $t/s$ - 
the result known from the constituent counting rules.\cite{ccr}

Presence of hard gluon exchange (Fig.1) indicates that one need sufficiently high energies 
to apply pQCD formalism here. However, since only one pion distribution amplitude is
involved for the direct process, if the photon attaches to the produced $q\overline q$ pair in Fig.1,
average virtuality of the gluon in question corresponds to
the one determining pion electromagnetic form factor at $Q^2\approx 20(35)$ GeV$^2$ scale, for
the asymptotic (Chernyak-Zhitnitsky) pion distribution amplitude
assuming CEBAF energy of 12 GeV, pion emission angle of 22$^o$ and $m_X$=2 GeV
(see Ref.\cite{acw} for details). 
Therefore one may hope to observe a single-gluon exchange, which is a higher-twist effect, in
inclusive photoproduction of pions even at CEBAF energies generally considered not high enough to
reach the perturbative QCD domain. Indications for presence of this direct pion 
production off a quark were obtained in $\pi N$ scattering 
(see Ref.\cite{owens} for references and discussion). Experimental observation of this mechanism 
at CEBAF would be very
important for our understanding of underlying mechanisms of exclusive and semi-exclusive reactions.

Testing scaling and duality in inclusive photoproduction of mesons requires coverage of the 
large-$x$ region, where the cross sections are rapidly falling as $x$ approaches 1. Therefore
such a uniquely designed high-lu\-mi\-no\-si\-ty machine as CEBAF can do an excellent job in these duality studies. 

\section{Elastic Form Factors and Nonforward Parton Densities}

Another way to relate inclusive and exclusive reactions may be realized through Nonforward Parton Densities introduced by Radyushkin
(NFPD).\cite{nfpd3} For related studies, see Refs.\cite{nfpd1,nfpd2}
 Recently Radyushkin~\cite{wacs} used the fact that NFPD satisfy 
certain sum rules giving elastic nucleon form factors (both vector and axial-vector) and produced an NFPD-based
model of elastic nucleon form factors.
Using standard parton densities measured in deep-inelastic lepton scattering and a Gaussian dependence of the
proton light-cone wave function on the transverse momentum, he obtained a good description 
of the proton Dirac form factor adjusting only one parameter: average transverse momentum 
carried by the quarks. The model, however has anomalous behavior in the low-$Q^2$ region because 
massless current (not constituent) quarks are considered in the calculations, therefore we will 
discuss here predictions for the form factors at $Q^2\geq 1$ GeV$^2$, except for the $Q^2=0$ point where
normalization of NFPD is fixed.

In order to analyze behavior of the Pauli form factor, $F_2(Q^2)$ one needs to construct a 
model of spin-flip NFPDs (${\mathcal K}^a(x,t)$). Assuming that spin-flip NFPD has similar dependence on the 
transverse momentum as spin-nonflip NFPD of Ref.\cite{wacs} and making an additional assumption
that the anomalous magnetic moment of the nucleon is determined only by valence quarks,
we are left only with one unknown ingredient of the model: the forward ($t=0$) limit of spin-flip NFPD,
$k_a(x)$.
\begin{eqnarray}
{\mathcal K}^a(x,t)=k_a(x) e^ {(1-x) t\over 4 x \lambda_k^2},
\end{eqnarray}
where $\lambda_k$ is a mass parameter.
The distribution $k_a(x)$ cannot be observed in deep-inelastic scattering, but may be accessed
in Deeply Virtual Compton Scattering.\cite{nfpd1,nfpd2,nfpd3} 

To build a model for
$k_a(x)$, I make a pQCD-based assumption that spin-flip deep-inelastic structure
 functions are suppressed at $x\to 1$ by an extra factor of $(1-x)^{2\delta h}$ compared 
 to spin-nonflip ones, where $\delta h$ 
 is the magnitude of helicity flip. It results in an extra factor of $(1-x)$ for spin-flip NFPD.
 Another constraint comes about in the $x\to 0$ limit if one recalls that spin-flip Regge amplitudes 
 are suppressed at high energies, giving  spin-flip NFPD an extra factor of  $x$. Based on these arguments, 
 I studied two choices of spin-flip NFPD: a) $k_a(x)=(1-x) f_a(x)$ and b) $k_a(x)= x(1-x) f_a(x)$.
 Adjusting the parameter $\lambda_k$ to fit the data on the proton $F_2/F_1$ ratio, the best description
 of experimental data from SLAC \cite{ffdata} is obtained with the choice (b) (at $\lambda_k^2=0.3$ GeV$^2$), 
 confirming the Regge argument in the low-$x$ region.

\begin{figure}[h]
\let\picnaturalsize=N
\def\picsize{2in}
\def\picfilenamea{bwQ2F2F1.eps}
\ifx\nopictures Y\else{\ifx\epsfloaded Y\else\input epsf \fi
\let\epsfloaded=Y
\centerline{
\ifx\picnaturalsize N\epsfxsize \picsize\fi \epsfbox{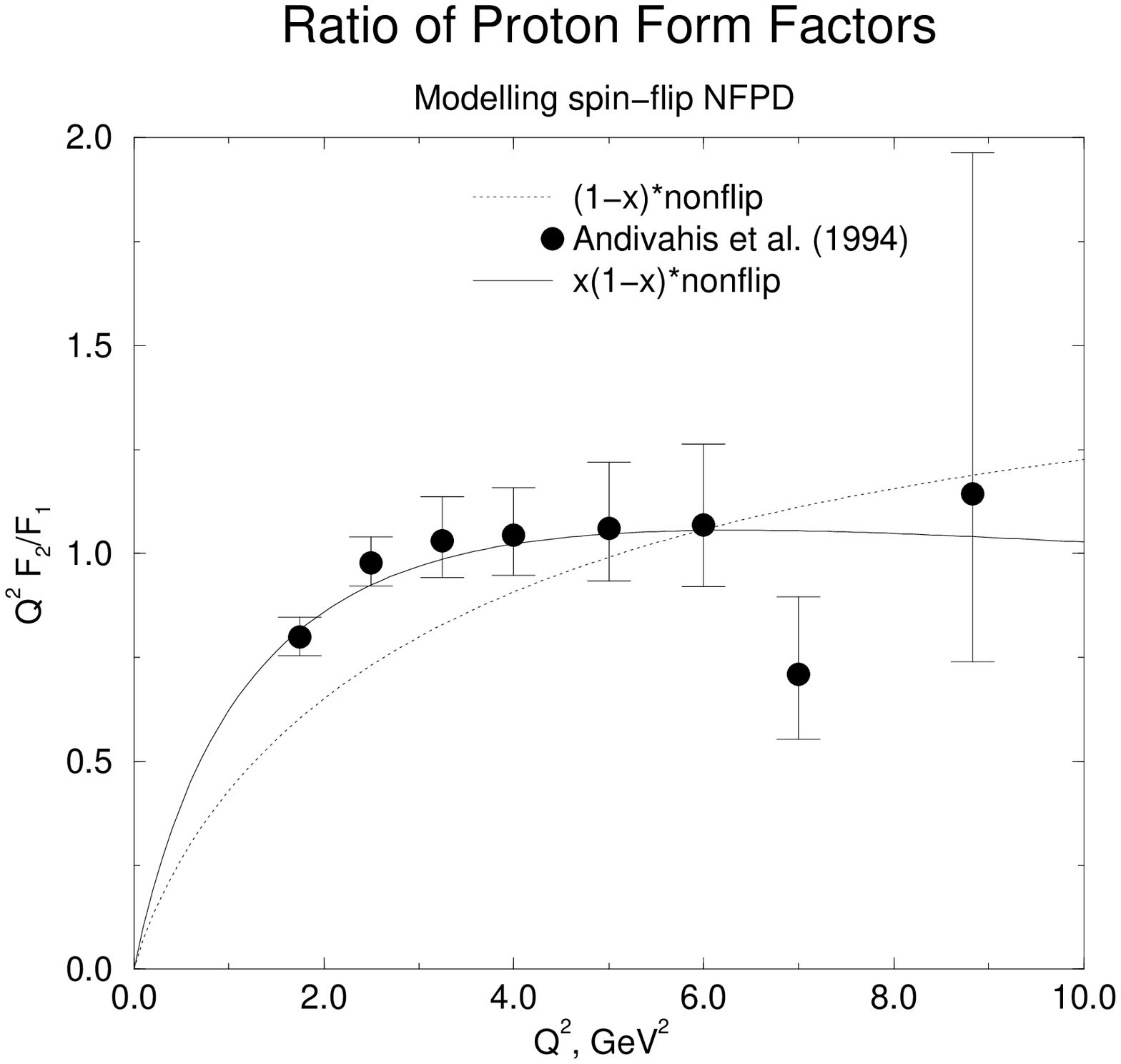}}}
\fi
\let\picnaturalsize=N
\def\picsize{2in}
\def\picfilenamea{bwGep.eps}
\ifx\nopictures Y\else{\ifx\epsfloaded Y\else\input epsf \fi
\let\epsfloaded=Y
\centerline{
\ifx\picnaturalsize N\epsfxsize \picsize\fi \epsfbox{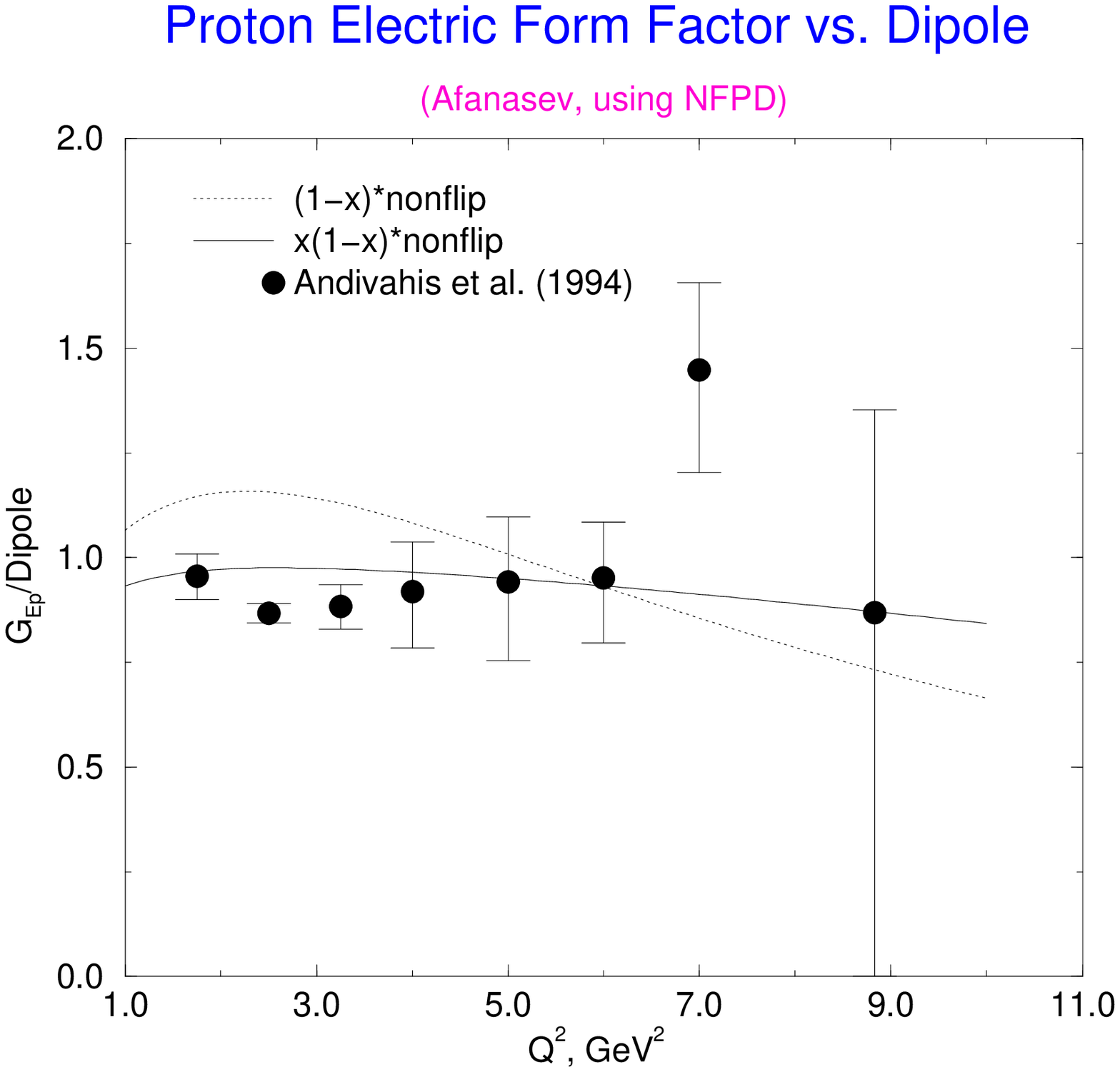}}}
\fi
\fcaption{Proton form factors from NFPD.}
\end{figure}

Both magnetic and electric proton form factors appear to be quite close to the phenomenological
dipole fit. Swapping up- and down-quark distributions yields neutron elastic form factors without
additional parameters (Fig.3) and with reasonable agreement with experiment.\cite{ffdata}
The available neutron form factor data provide even stronger evidence in favor of additional suppression
of $k_a(x)$ at low $x$. Predictions for the electric neutron form factor are also presented in Fig.3.
One can see that the predictions for the elastic nucleon form factors are very sensitive to the models of
nucleon NFPD, namely, $\mathcal F$ and $\mathcal K$. Therefore Jefferon Lab experimental program of precise
measurements of nucleon form factor is very important for obtaining constraints on NFPD.

\begin{figure}[h]
\let\picnaturalsize=N
\def\picsize{2in}
\def\picfilenamea{bwGmn.eps}
\ifx\nopictures Y\else{\ifx\epsfloaded Y\else\input epsf \fi
\let\epsfloaded=Y
\centerline{
\ifx\picnaturalsize N\epsfxsize \picsize\fi \epsfbox{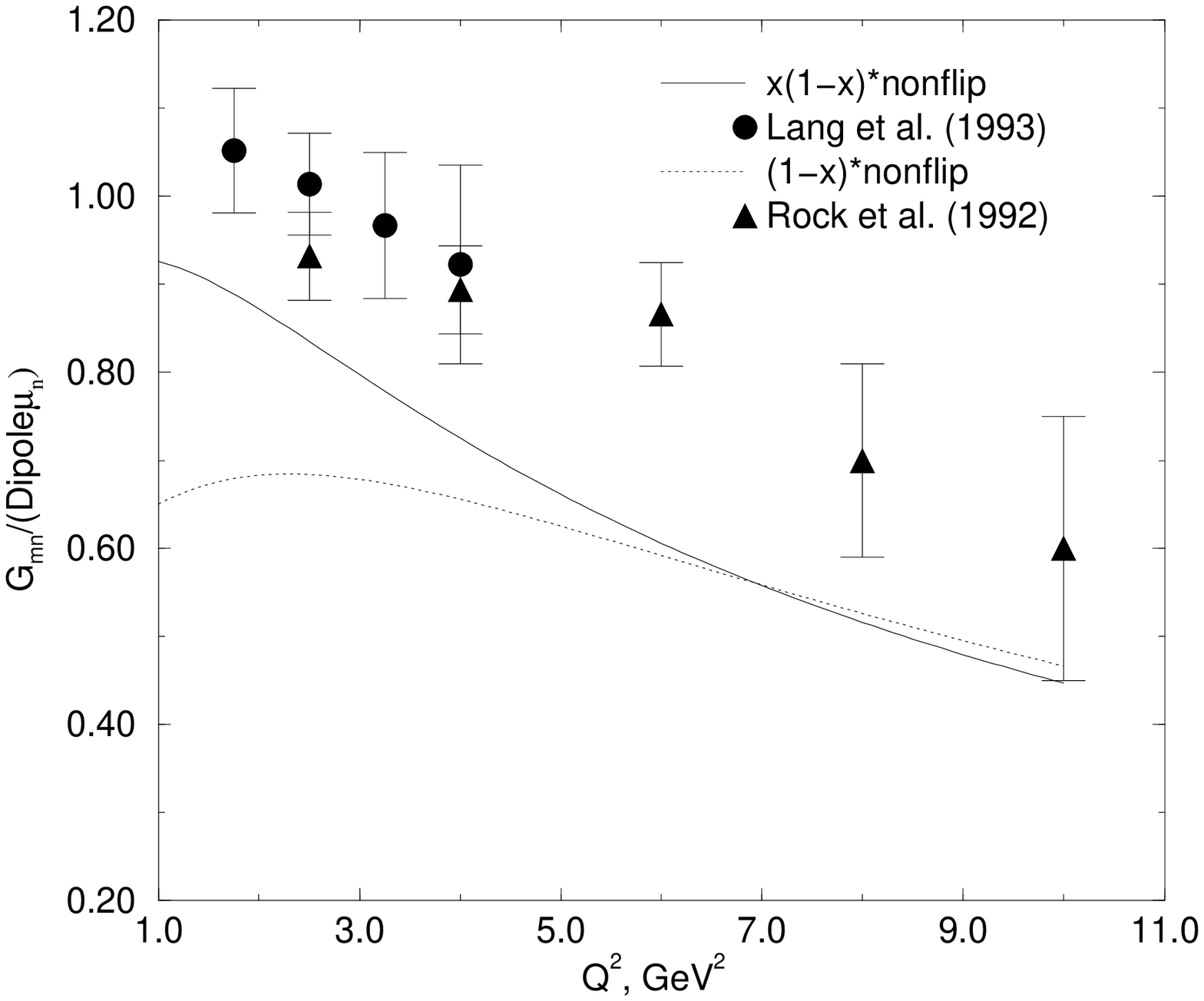}}}
\fi
\let\picnaturalsize=N
\def\picsize{2in}
\def\picfilenamea{bwGen.eps}
\ifx\nopictures Y\else{\ifx\epsfloaded Y\else\input epsf \fi
\let\epsfloaded=Y
\centerline{
\ifx\picnaturalsize N\epsfxsize \picsize\fi \epsfbox{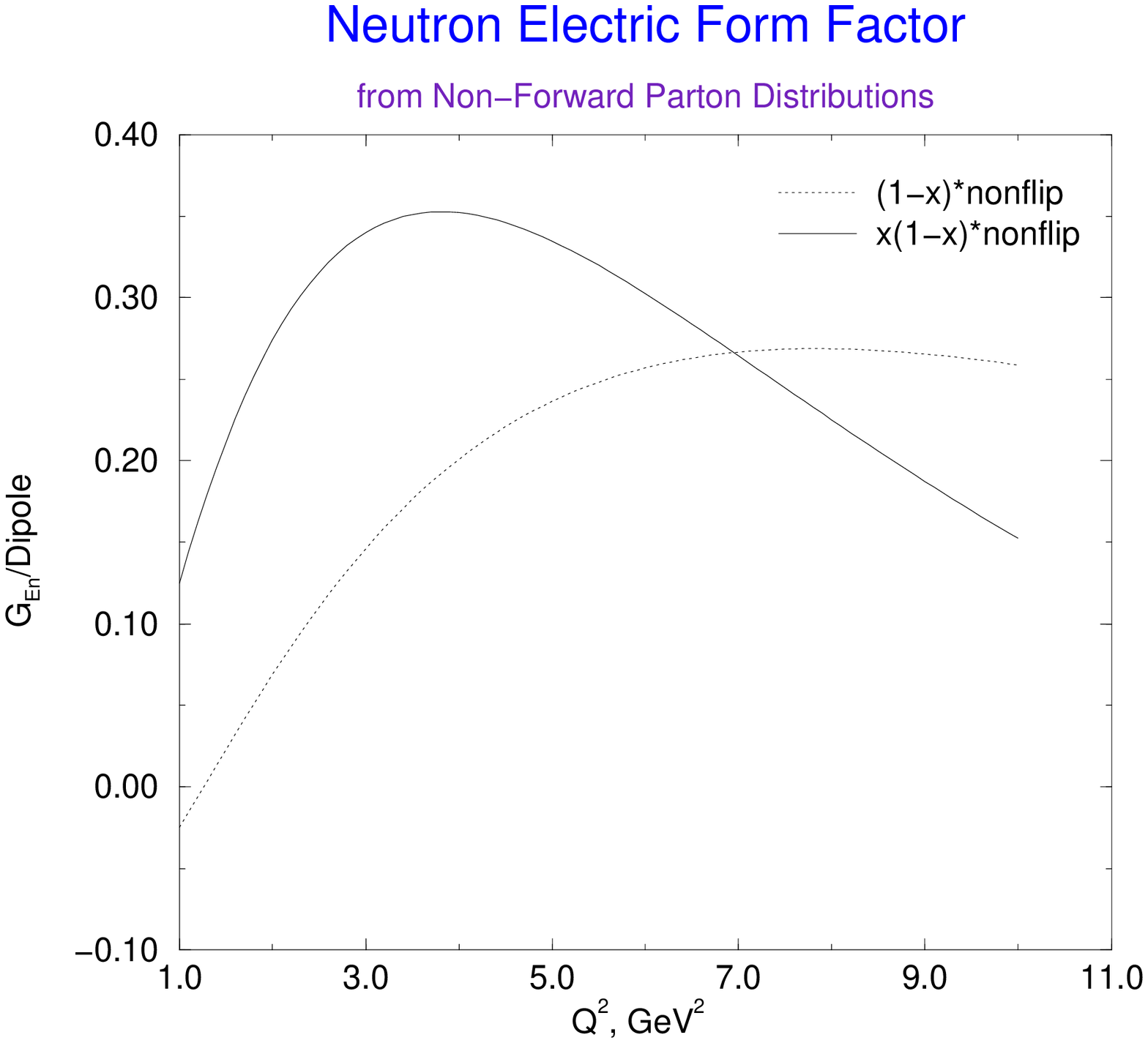}}}
\fi
\fcaption{Neutron form factors from NFPD with the same parameters as in Fig.2.}
\end{figure}

The same method also allows one to compute axial-vector form factor of the nucleon, $F_A(Q^2)$. 
Assuming the same value of the transverse momentum parameter $\lambda^2=0.7$ GeV$^2$ for polarized 
quarks as for unpolarized ones, and using a
phenomenological parametrization of polarized parton distributions from deep-inelastic experiments,
 I found $F_A(Q^2)$ close to the dipole form
with a dipole mass $M_A\approx 1.1$ GeV, in reasonable agreement with neutrino-nucleon elastic 
scattering data.\cite{axial}

Using NFPD for the pion in the form 
${\mathcal F}^\pi_a(x,t)\propto{(1-x)\over \sqrt{x}} \exp[{(1-x) t\over 4 x \lambda^2}]$, where the $t\to 0$ 
limit is justified by the analysis\cite{smrs} of all available Drell-Yan and prompt photon $\pi N$ data, 
and taking $\lambda^2=0.7$ GeV$^2$,
 I obtained the result for
the pion elastic form  factor shown in Fig.4. It is close to the one predicted in QCD Sum Rules,\cite{nerad}
but has a different shape due to different $x$-dependence at low $x$, ($\propto 1/\sqrt{x}$ as required by 
Regge constraint vs. $\propto x$ in the QCD Sum Rules). Both models predict deviations from the constituent counting
rules.\cite{ccr} 

\begin{figure}[h]
\let\picnaturalsize=N
\def\picsize{2in}
\def\picfilenamea{pionff.eps}
\ifx\nopictures Y\else{\ifx\epsfloaded Y\else\input epsf \fi
\let\epsfloaded=Y
\centerline{
\ifx\picnaturalsize N\epsfxsize \picsize\fi \epsfbox{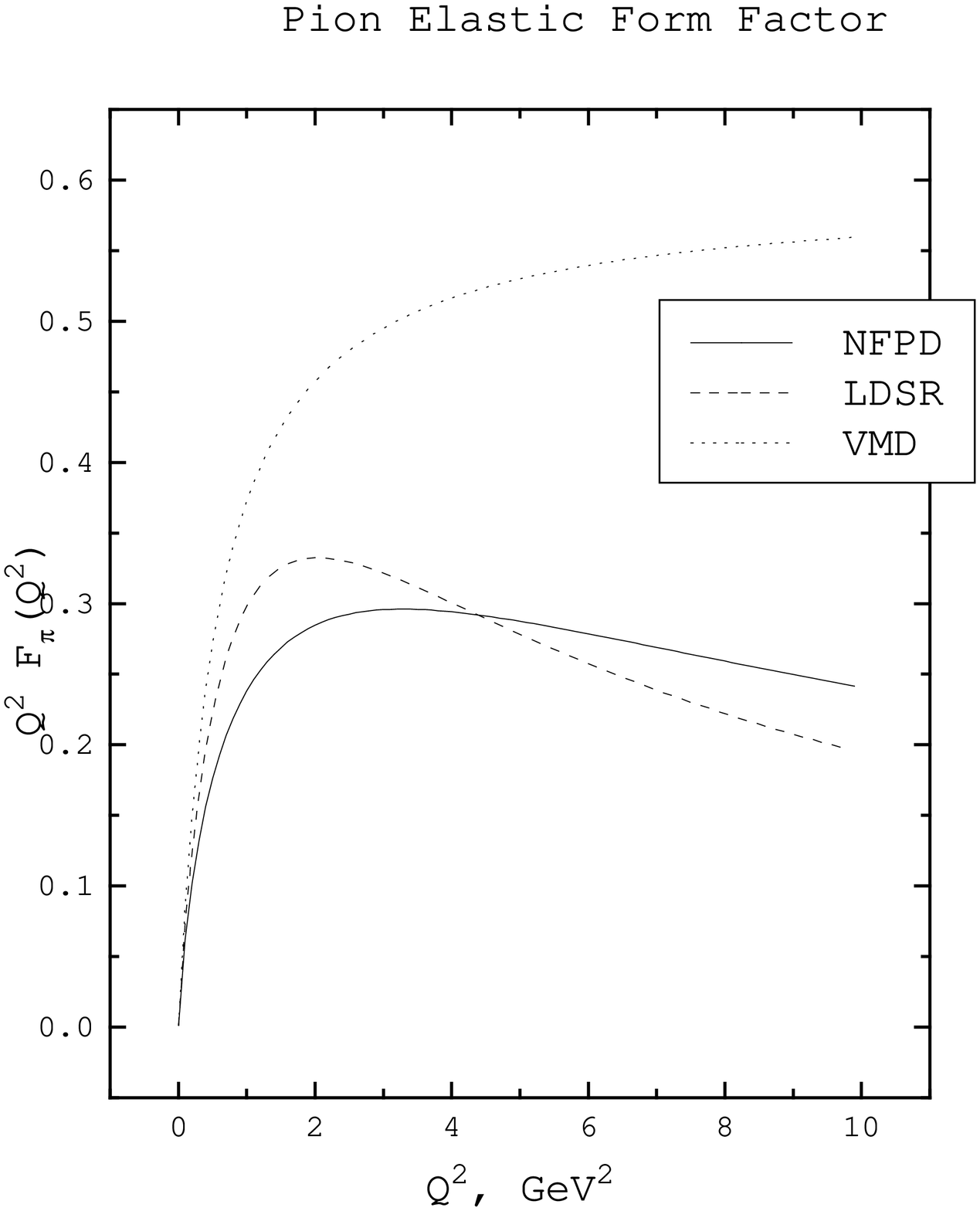}}}
\fi
\fcaption{Elastic pion form factor from NFPD with $\lambda^2=0.7$ GeV$^2$. 
Curves for the $\rho$-dominance (VMD) and
Local Duality Sum Rule (LDSR) Ref.$^9$ are shown for comparison.}
\end{figure}

With this model of pion NFPD, I calculated cross sections for the pion Compton scattering,
$\gamma\pi\to\gamma\pi$
(the method of calculation was similar to the one used in Ref.\cite{wacs}). The results, same for the 
neutral and charge pions (since both initial and final photons are assumed to couple to the same quark, 
which is justified at high energies), are shown in Fig. 5. One can see deviation from constituent scaling behavior
as a function of the c.m. scattering angle ($\theta_{c.m.}$) with a general trend: the smaller is
$\theta_{c.m.}$,
the slower the cross section falls with $s$. Only around $\theta_{c.m.}$= 45$^o$, the cross section
falls approximately as $1/s^4$, as predicted by perturbative QCD,\cite{ccr} but this is accidental since in our 
model the reaction proceeds through the overlap of initial and final pion wave functions (described by NFPD),
not caused by hard gluon exchange.

\begin{figure}[h]
\let\picnaturalsize=N
\def\picsize{2.5in}
\def\picfilenamea{picomp.eps}
\ifx\nopictures Y\else{\ifx\epsfloaded Y\else\input epsf \fi
\let\epsfloaded=Y
\centerline{
\ifx\picnaturalsize N\epsfxsize \picsize\fi \epsfbox{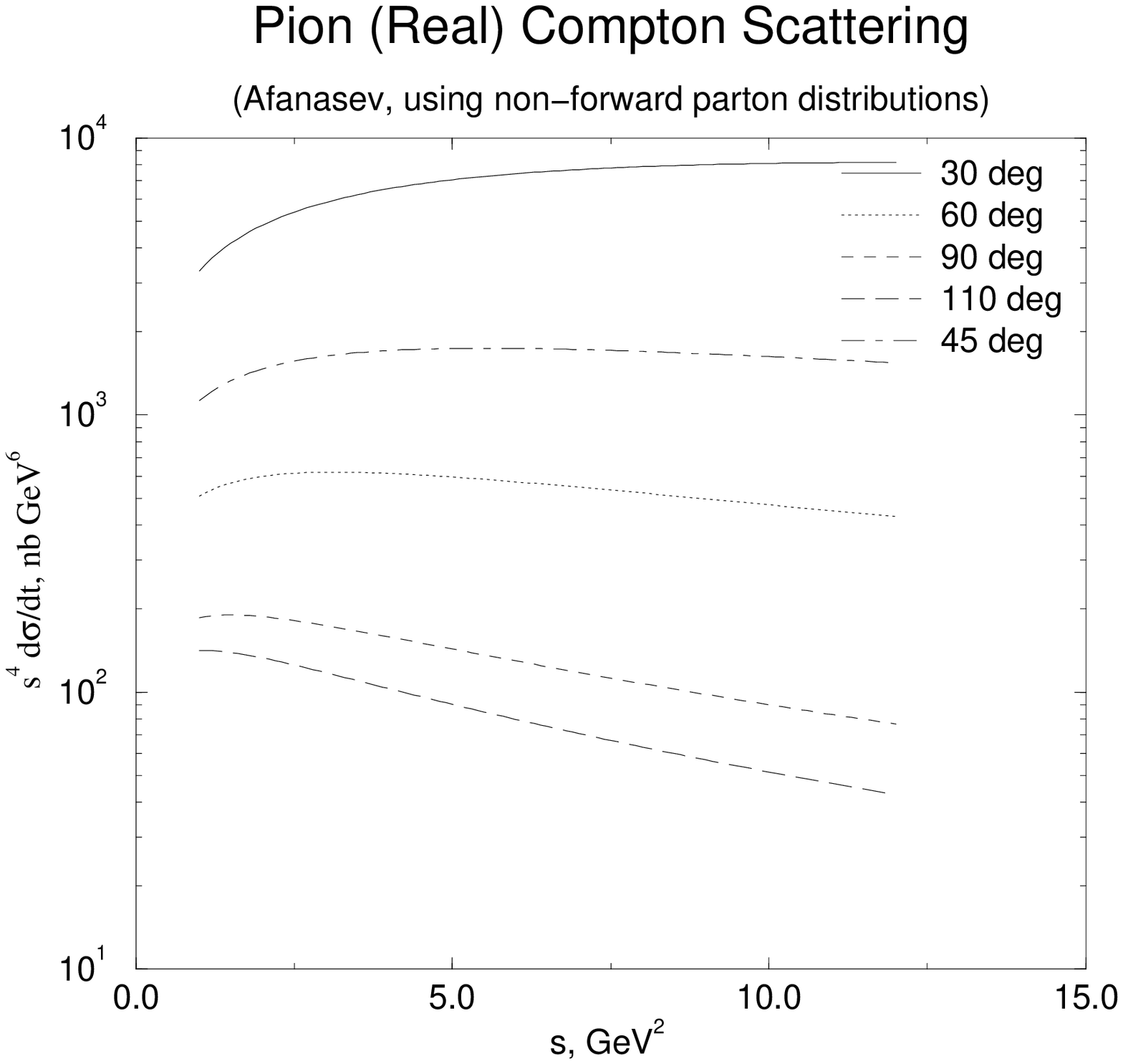}}}
\fi
\fcaption{Deviation from the constituent counting rule for pion Compton scattering as a function 
of photon c.m. scattering angle, obtained using pion NFPD with the parameters as in Fig.4.}
\end{figure}

Both pion form factor and pion Compton scattering may be studied on a fixed nucleon target
at CEBAF in a peripheral process such that a pion from
a meson cloud surrounding the nucleon absorbs the most of the momentum of the
incident (either real or virtual) photon. Pion form factor measurements are being
analyzed after the first (lower energy) run of JLab-E-93-021.\cite{mack}
Pion (both real and virtual) Compton scattering may be probed at CEBAF with radiative pion photoproduction, 
$\gamma p\to\gamma\pi N$, taking advantage of the third-arm photon spectrometer to be installed at Hall A.

Future experiments at Jefferson Lab will provide precise tests of deviation from constituent 
scaling behavior of cross sections of exclusive electromagnetic processes involving
pions and nucleons.

\section{Acknowledgements}

Collaboration with C.E. Carlson, A.V. Radyushkin and C. Wahlquist is gratefully acknowledged.
This work was partially supported by the U.S. Department of Energy under
contract DE--AC05--84ER40150.

\end{document}